\documentclass[12pt]{article}
\usepackage{amssymb,amscd,array}
\catcode `\@=11
\@addtoreset{equation}{subsection}

\def\harr#1#2{\smash{\mathop{\hbox to .5in{\rightarrowfill}}
\limits^{\scriptstyle#1}_{\scriptstyle#2}}}

\def\harrl#1#2{\smash{\mathop{\hbox to .5in{\leftarrowfill}}
\limits^{\scriptstyle#1}_{\scriptstyle#2}}}

\def\qed{\blacksquare}
\newcommand{\be}{\begin{equation}}
\newcommand{\ee}{\end{equation}}
\newcommand{\bea}{\begin{eqnarray}}
\newcommand{\eea}{\end{eqnarray}}
\newcommand{\R}{\mathbb{R}}
\newcommand{\N}{\mathbb{N}}
\newcommand{\C}{\mathbb{C}}
\newtheorem{thm}{Theorem}[section]
\newtheorem{rem}[thm]{Remark}

\newtheorem{prop}[thm]{Proposition}
\begin{document}
\begin{titlepage}
\begin{center}
{\bf \Large{Wess-Zumino Model in the Causal Approach\\}}
\end{center}
\vskip 1.0truecm
\centerline{D. R. Grigore
\footnote{e-mail: grigore@theor1.theory.nipne.ro, grigore@theory.nipne.ro}}
\vskip5mm
\centerline{Dept. of Theor. Phys., Inst. Atomic Phys.}
\centerline{Bucharest-M\u agurele, P. O. Box MG 6, ROM\^ANIA}
\vskip 2cm
\bigskip \nopagebreak
\begin{abstract}
\noindent
The Wess-Zumino model is analysed in the framework of the causal approach of
Epstein-Glaser. The condition of invariance with respect to supersymmetry
transformations is similar to the gauge invariance in the Z\"urich formulation.
We prove that this invariance condition can be implemented in all orders of
perturbation theory, i.e. the anomalies are absent in all orders. This result
is of purely algebraic nature. We work consistently in the quantum framework
based on Bogoliubov axioms of perturbation theory so no Grassmann variables
are necessary.
\end{abstract}
\end{titlepage}

\section{Introduction}

The causal approach to renormalization theory of by Epstein and Glaser
\cite{EG1}, \cite{Gl} seems to be the most convenient way to understand
renormalization theory at the fundamental level. It is also extremely useful
for purely computational aspects. In this paper we will prove that
supersymmetric theories can be also studied in a completely rigorous way in
this framework. We will analyse the simplest supersymmetric model, namely the
Wess-Zumino model \cite{WZ}, \cite{IZ}. We do not use in this paper the
superfield formulation \cite{SS}, \cite{WB}, \cite{We}. We prefer to formulate
this model working directly in the quantum framework: we consider in the Fock
space of the model (generated by a scalar, a pseudo-scalar and a Majorana
quantum free fields of the same positive mass) and construct the chronological
products verifying Bogoliubov axioms. We can define in this Fock space the
supersymmetric current and the supercharge; they are only the linear
contributions of the usual expressions appearing in the literature. Then we
impose the condition of supersymmetry invariance at the quantum level in close
analogy to the condition of gauge invariance adopted by the Z\"urich group for
gauge theories \cite{DHKS1}, \cite{DHKS2}; the physical meaning of this
condition is the invariance of the $S$-matrix with respect to supersymmetric
transformations in the adiabatic limit. In the next Section we give the
essential points concerning the perturbation theory in the sense of Bogoliubov
and Epstein-Glaser; for more details see \cite{qed} and literature cited there.
In Section \ref{wz} we define the Wess-Zumino model in this framework and in
Section \ref{ward} we prove that supersymmetry invariance can be implemented to
all orders of perturbation theory by a purely algebraic procedure of
distribution splitting.
\newpage

\section{Perturbation Theory in the Causal Approach\label{pert}}

\subsection{Bogoliubov Axioms}{\label{bogoliubov}}

Let us remind briefly the main ideas of Epstein-Glaser-Scharf approach.
According to Bogoliubov and Shirkov, the $S$-matrix is constructed inductively
order by order as a formal series of operator valued distributions:
\be
S(g) = 1 + \sum_{n=1}^\infty{i^{n}\over n!}\int_{\R^{4n}} dx_{1}\cdots dx_{n}\,
T(x_{1},\cdots, x_{n}) g(x_{1})\cdots g(x_{n}),
\label{S}
\ee
where
$g(x)$
is a tempered test function in the Minkowski space 
$\R^{4}$
that switches the interaction and
$T(X) \equiv T(x_{1},\cdots, x_{n})$
are operator-valued distributions acting in the Hilbert space 
${\cal H}$
generated by some collection of free fields. These operator-valued
distributions, which are called {\it chronological products} should verify some
properties which can be argued starting from {\it Bogoliubov axioms}.  We give
here the set of axioms imposed on the chronological products following 
\cite{Gl}.
\begin{itemize}
\item
Domain: There is a common dense domain of definition
$D_{0} \in {\cal F}$
for all chronological products.
\item
Symmetry:
\be
T(x_{\pi(1)},\cdots x_{\pi(n)}) = T(x_{1},\cdots x_{n}), \quad \forall
\pi \in {\cal P}_{n}.
\label{sym}
\ee
\item
Poincar\'e invariance: There exists in the Fock space of the model an unitary
representation 
$(a, A) \mapsto U_{a, A}$ 
of the group $inSL(2,\C)$ 
(the universal covering group of the proper orthochronous Poincar\'e group
${\cal P}^{\uparrow}_{+}$ 
- see \cite{Va} for notations) such that:
\be
U_{a, A} T(x_{1},\cdots, x_{n}) U^{-1}_{a, A} =
T(\delta(A)\cdot x_{1}+a,\cdots, \delta(A)\cdot x_{n}+a), 
\quad \forall A \in SL(2,\C), \forall a \in \R^{4}
\label{invariance}
\ee
where
$SL(2,\C) \ni A \delta(A) \in {\cal P}^{\uparrow}_{+}$
is the covering map.  In particular, {\it translation invariance} is essential
for implementing Epstein-Glaser scheme of renormalization.

Sometimes it is possible to supplement this axiom by invariance properties with
respect to space-time inversions, charge conjugation or invariance with respect
to some global group of transformations (continuous or discrete). In this paper
we will impose the invariance with respect to supersymmetry transformations.
\item
Causality. We use the standard notations:
$
V^{\pm} \equiv \{x \in \R^{4} \vert \quad x^{2} > 0, \quad 
\mbox{sign}(x_{0}) = \pm\}
$
for the upper (lower) lightcone and
$\overline{V^{\pm}}$
for their closures. If
$X \equiv \{x_{1},\cdots, x_{m}\} \in \R^{4m}$
and
$Y \equiv \{y_{1},\cdots, y_{n}\} \in \R^{4m}$
are such that
$
x_{i} - y_{j} \not\in \overline{V^{-}}, \quad \forall i=1,\dots,m,\quad
j=1,\dots,n
$
we use the notation
$X \geq Y.$
If
$
x_{i} - y_{j} \not\in \overline{V^{+}} \cup \overline{V^{-}}, 
\quad \forall i=1,\dots,m,\quad
j=1,\dots,n
$
we use the notations:
$X \sim Y.$
Then the causality
axiom writes as follows:
\be
T(X_{1}X_{2}) = T(X_{1}) T(X_{2}), 
\quad \forall X_{1} \geq X_{2}.
\label{causality}
\ee
\item
Unitarity: We define the expressions
\be
(-1)^{|X|} \bar{T}(X) \equiv \sum_{r=1}^{|X|} (-1)^{r} 
\sum_{X_{1},\cdots,X_{r} \in Part(X)} T(X_{1})\cdots T(X_{r});
\label{antichrono}
\ee
One calls the operator-valued distributions
$\bar{T(X)}$
{\it anti-chronological products}. Then the unitarity axiom is:
\be
\bar{T}(X) = T(X)^{*}, \quad \forall X.
\label{unitarity}
\ee
\end{itemize}
\subsection{Epstein-Glaser Induction}{\label{EG}}

In this Subsection we summarize the steps of the inductive construction of
Epstein and Glaser \cite{EG1}. The main point is a careful formulation of the
induction hypothesis. So, we suppose that we have the {\it interaction 
Lagrangian}
$T(x)$
given by a sum of Wick monomials acting in a certain Fock space. We make the
simplifying assumption (valid for the Wess-Zumino model) that {\bf no
derivative of the fields appear} in the Wick monomials composing 
$T(x)$.
Moreover, we require the following properties:
\be
U_{a,A} T(x) U^{-1}_{a,A} = T(\delta(A)\cdot x+a),
\quad \forall A \in SL(2,\C),
\label{inv1}
\ee
\be
\left[T(x), T(y)\right] = 0, \quad \forall x,y \in \R^{4} \quad
s.t. \quad x \sim y,
\label{causality1}
\ee
and
\be
T(x)^{*} = T(x).
\label{unitarity1}
\ee

Usually, these requirements are supplemented by covariance with respect to some
discrete symmetries like space-time inversions, charge conjugations or global
invariance with respect to some Lie group of symmetry. In this paper we will
add supersymmetry invariance (see the next Section)..

Let us define the {\it degree} of a Wick monomial 
$deg(W)$
by assigning to every integer spin field factor and every derivative the value
$1$, for every half-integer spin field factor the value $3/2$ and summing over
all factors. We consider that the interaction Lagrangian has the canonical
dimension 
$\leq 4$.

We suppose that we have constructed the chronological products 
$T(X), \quad |X| \leq n - 1$
having the following properties: (\ref{sym})- (\ref{causality}) and
(\ref{unitarity}). We add to the induction hypothesis the following {\it Wick
expansion} property:
\be
T(X) = \sum_{i} t_{i}(X) W_{i}(X), \quad |X| \leq n - 1
\label{wick}
\ee
where
$W_{i}(X)$
are basis of linearly independent Wick monomials {\bf without derivatives on
the fields} and
$t_{i}(X)$
are numerical distributions; they are called {\it renormalized Feynman
amplitudes} and are Poincar\'e covariant. Finally, the following limitations
is included in the induction hypothesis:
\be
\omega(t_{i}) + deg(W_{i}) \leq 4, \quad \forall i
\label{deg-t}
\ee
where by
$\omega(t)$
we mean the order of singularity of the distribution $t$ (see \cite{Sc1} for
the definition).

Let us note that in this case we also have:
\be
[T(X_{1}), T(X_{2})] = 0, \quad {\rm if}  \quad X_{1} \sim X_{2}, \quad
|X_{1}| + |X_{2}| \leq n - 1.
\label{commute}
\ee

We want to construct the distribution-valued operators
$T(X), \quad |X| = n$
such that the induction hypothesis stays true.

Here are the main steps of the induction proof.
\begin{enumerate}
\item
One constructs from
$T(X), \quad |X| \leq n - 1$
the expressions
$\bar{T}(X), \quad |X| \leq n - 1$
according to (\ref{antichrono}).
\item
One defines the expressions:
\be
A'(x_{1},\dots,x_{n-1};x_{n}) \equiv 
\sum_{\stackrel{X_{1},X_{2} \in Part(X)}{X_{2} \not= \emptyset, 
x_{n} \in X_{1}}}
(-1)^{|X_{2}|} T(X_{1}) \bar{T}(X_{2}),
\label{A1}
\ee
\be
R'(x_{1},\dots,x_{n-1};x_{n}) \equiv 
\sum_{\stackrel{X_{1},X_{2} \in Part(X)}{X_{2} \not= \emptyset, 
x_{n} \in X_{1}}}
(-1)^{|X_{2}|} \bar{T}(X_{2}) T(X_{1})
\label{R1}
\ee
and
\be
D(x_{1},\dots,x_{n-1};x_{n}) \equiv A'(x_{1},\dots,x_{n-1};x_{n}) -
R'(x_{1},\dots,x_{n-1};x_{n}).
\label{D}
\ee

The one can prove that we have the causal support property:
\be
supp(D(X)) \subset \Gamma^{+}(x_{n}) \cup \Gamma^{-}(x_{n})
\ee
where we use standard notations:
\be
\Gamma^{\pm}(x_{n}) \equiv \{ (x_{1},\dots,x_{n}) \in (\R^{4})^{n} |
x_{i} - x_{n} \in V^{\pm} , \quad \forall i = 1, \dots, n-1\}
\ee
\item
The distribution 
$D(X)$
can be written as a sum
\be
D(X) = \sum_{i} d_{i}(X) W_{i}(X)
\ee
where
$d_{i}(X)$
are numerical distributions with causal support i.e
\be
supp(d_{i}(X)) \subset \Gamma^{+}(x_{n}) \cup \Gamma^{-}(x_{n})
\ee
and Poincar\'e covariant. Finally, the following limitations are valid:
\be
\omega(d_{i}) + deg(W_{i}) \leq 4, \quad \forall i.
\label{deg-d}
\ee

Let us note that in theories with derivatives it is much more difficult to
extract the properties of the numerical distributions
$d_{i}$
from the corresponding properties of the operatorial distribution
$D(X)$:
one has a supplementary induction hypothesis concerning the Wick submonomials
\cite{EG1}, \cite{DF}.
\item
Now we have the following result from \cite{DHKS2}, \cite{Sc1}: Let $d$ be a
$SL(2,\C)$-covariant
distribution with causal support. Then, there exists a causal splitting
\be
d = a - r, \quad supp(a) \subset \Gamma^{+}(x_{n}), \quad 
supp(r) \subset \Gamma^{-}(x_{n})
\ee
which is also
$SL(2,\C)$-covariant and such that 
\be
\omega(a) \leq \omega(d), \quad \omega(r) \leq \omega(d).
\ee

So, there exists a 
$SL(2,\C)$-covariant causal splitting:
\be
D(x_{1},\dots,x_{n-1};x_{n}) =
A(x_{1},\dots,x_{n-1};x_{n}) - R(x_{1},\dots,x_{n-1};x_{n})
\label{decD}
\ee
with
$supp(A(x_{1},\dots,x_{n-1};x_{n})) \subset \Gamma^{+}(x_{n})$
and
$supp(R(x_{1},\dots,x_{n-1};x_{n})) \subset \Gamma^{-}(x_{n})$.

For that reason, the expressions
$A(X)$ 
and
$R(X)$
are called {\it advanced} (resp.. {\it retarded}) products.
\item
One can prove that the following relation is true
\be
D(X)^{*} = (-1)^{n-1} D(X).
\label{conj}
\ee
As a consequence, the causal splitting obtained above can be chosen such that
\be
A(X)^{*} = (-1)^{n-1} A(X).
\ee

This can be done by the redefinition:
\be
A(X) \rightarrow {1\over 2} \left[ A(X) + (-1)^{n-1} A(X)^{*} \right]
\label{redef}
\ee
which does not affect the support property.
\item
Let us define
\be
T'(X) \equiv A(X) - A'(X) = R(X) - R'(X).
\ee
Then these expressions satisfy the 
$SL(2,\C)$-covariance, 
causality and unitarity conditions (\ref{invariance}), (\ref{causality}),
(\ref{unitarity}) and Wick expansion property. If we substitute:
\be
T(x_{1},\cdots, x_{n}) \rightarrow {1 \over n!}
\sum_{\pi} T'(x_{\pi(1)},\cdots, x_{\pi(n)})
\label{symmetrization}
\ee
where the sum runs over all permutations of the numbers
$\{1, \dots, n\}$
then we also have the symmetry axiom (\ref{sym}). 
\end{enumerate}

The solution of the renormalization problem is not unique. If all chronological
products up to order 
$n-1$
are determined, then the non-uniqueness in order $n$
is given by the possibility of adding to the distributions
$T(X), \quad |X| = n$
some finite renormalizations (quasi-local operators in the terminology of
\cite{BS2})
$N(X)$.
\newpage

\section{Wess-Zumino Model\label{wz}}

\subsection{The Definition of the Model}

In this Subsection we define the Wess-Zumino model in the framework of
Bogoliubov axioms presented above. We considers the Hilbert space
$\cal H$
endowed with the scalar product
$<\cdot,\cdot>$
and generated by applying on the vacuum 
$\Omega$ 
the following free fields: the scalar field
$A(x)$,
the pseudo-scalar field
$B(x)$
and the Majorana field
$\psi(x)$.
These fields are assumed to have the same mass
$m > 0$.

To describe the Majorana field we need Dirac matrices 
$\gamma^{\mu}, \quad \mu = 0,\dots,3$ for which we prefer the chiral
representation:
\be
\gamma_{0} = \left( \matrix{ 0 & 1 \cr 1 & 0} \right), \quad
\gamma_{i} = \left( \matrix{ 0 & - \sigma_{i} \cr \sigma_{i} & 0} \right),
\quad i = 1,2,3;
\ee
here
$\sigma_{i}, \quad i = 1,2,3$
are the Pauli matrices. This is a representations in which the matrix
$\gamma_{5} \equiv i \gamma_{0}\gamma_{1}\gamma_{2}\gamma_{3}$
is diagonal:
\be
\gamma_{5} = \left( \matrix{ 1 & 0 \cr 0 & -1} \right).
\ee

If 
$u \in \C^{4}$ 
is a spinor considered as a column vector then we define
$\bar{u} \equiv u^{*} \gamma_{0}$
considered as a line vector.

The fields considered in our model are determined by the following properties:
\begin{itemize}
\item
Equations of motion:
\be
(\partial^{2} + m^{2}) A(x) = 0, \quad 
(\partial^{2} + m^{2}) B(x) = 0, \quad 
(i \gamma \cdot \partial - m) \psi(x) = 0.
\label{equ}
\ee
\item
Canonical (anti)commutation relations:
\bea
\left[ A(x), A(y) \right] = D_{m}(x-y) \times {\bf 1}, \quad
\left[ B(x), B(y) \right] = D_{m}(x-y) \times {\bf 1}
\nonumber \\
\{\psi_{\alpha}(x), \psi_{\beta}(y)\} = 
\left(S_{m}(x-y) C\right)_{\alpha\beta} \times {\bf 1}.
\eea
and all other (anti)commutators are null; here 
$C = \gamma_{0} \gamma_{2}$ 
is the charge conjugation matrix and
$S_{m}(x), \quad m \geq 0$
is a $4 \times 4$ matrix given by:
\be
S_{m}(x) \equiv (i \gamma \cdot \partial + m) D_{m}(x).
\label{Sm}
\ee
\item
$SL(2,\C)$-covariance:
\bea
U_{a,A} A(x) U^{-1}_{a,A} = A(\delta(A) \cdot x + a), \quad
U_{a,A} B(x) U^{-1}_{a,A} = B(\delta(A) \cdot x + a),
\nonumber \\
U_{a,A} \psi(x) U^{-1}_{a,A} = S(A^{-1}) \psi(\delta(A) \cdot x + a).
\nonumber \\
U_{a,A} \Omega = \Omega;
\label{poincare}
\eea
here
$\delta: SL(2,\C) \rightarrow {\cal L}^{\uparrow}_{+}$
is the covering map and
\be
S(A) \equiv \left( \matrix{ A & 0 \cr 0 & (A^{-1})^{*}} \right).
\ee
\item
Space-time covariance:
\bea
U_{I_{s}} A(x) U_{I_{s}}^{-1} = A(I_{s}\cdot x),
U_{I_{s}} B(x) U_{I_{s}}^{-1} = - B(I_{s}\cdot x),
U_{I_{s}} \psi(x) U_{I_{s}}^{-1} =i \gamma_{0} \psi(I_{s}\cdot x).
\label{spatial}
\eea
\bea
U_{I_{t}} A(x) U_{I_{t}}^{-1} = A(I_{t}\cdot x),
U_{I_{t}} B(x) U_{I_{t}}^{-1} = B(I_{t}\cdot x),
U_{I_{t}} \psi(x) U_{I_{t}}^{-1} = C^{-1} \gamma_{5} \psi(I_{t}\cdot x).
\label{temporal}
\eea
The space-time inversion is:
$U_{I_{st}} \equiv U_{I_{s}}~U_{I_{t}}$.
\item                        
Hermitian conjugation properties:
\be
A_{a\mu}(x)^{*} = A_{a\mu}(x), \quad
B(x)^{*} = B(x), \quad
\psi(x)^{c} = \psi(x);
\label{conjugate}
\ee
where $*$ is the conjugation with respect to the scalar product
$<\cdot,\cdot>$
and the definition of the charge conjugate of the spinor 
$u \in \C^{4}$
is:
\be
u^{c} \equiv C \bar{u}^{T}.
\ee
\item
Charge conjugation invariance:
\bea
U_{C} A(x) U_{C}^{-1} = A(x),
\quad
U_{C} B(x) U_{C}^{-1} = B(x),
\quad
U_{C} \psi(x) U_{C}^{-1} = C \bar{\psi(x)}^{T} = \psi(x).
\label{charge}
\eea
\item
Moreover, we suppose that these operators are leaving the vacuum invariant:
\be
U_{a,A} \Omega = \Omega, \quad 
U_{I_{s}} \Omega = \Omega, \quad 
U_{I_{t}} \Omega = \Omega, \quad 
U_{C} \Omega = \Omega.
\label{inv-vacuum}
\ee
\end{itemize}

\begin{rem}
One can prove that the operators
$U_{a,A}, \quad U_{I_{s}}$
and
$U_{I_{t}}$
are realizing a projective representation of the Poincar\'e group i.e. they
have suitable commutation properties. Also the charge conjugation operator
commutes with these operators.  As it is known, there is some freedom in
choosing some phases in the definitions of the spatial and temporal inversions;
we have made the convenient choice which ensures this commutativity property.
\label{projective}
\end{rem}

In this Fock space we can define the spinorial operator:
$J^{\mu}_{\alpha}(x) \equiv J^{\mu}_{\alpha}(x)_{\alpha = 1}^{4}$
called {\it supercurrent} according to the formula:
\be
J^{\mu} \equiv 
:\partial_{\nu}A  \gamma^{\nu} \gamma^{\mu}\psi_{\beta}:
+ i :\partial_{\nu}B  \gamma_{5} \gamma^{\nu} \gamma^{\mu}\psi_{\beta}:
+ im :A \gamma^{\mu}\psi_{\beta}:
+ i :B \gamma_{5} \gamma^{\mu}\psi_{\beta}:
\ee
where the interpretation of this operator as a column vector with four
components is obvious. Then we have by direct computation, using the equations
of motion, the following {\it conservation law}:
\be
\partial_{\mu} J^{\mu} = 0.
\ee

Moreover, the supercurrent, considered as a spinor, is charge conjugation
invariant:
\be
\left( J^{\mu}\right)^{c} = J^{\mu}.
\ee

One can define formally the {\it supercharges} as a four-component operator 
according to:
\be
Q_{\alpha} = \int_{\R^{3}} d^{3}x  J^{0}_{\alpha}(x).
\label{super1}
\ee

To avoid problems connected with the existence of the integral, it is better to
work in momentum space. One has the standard expressions of the free fields
considered in the model:
\be
A(x) \equiv {1\over (2\pi)^{3/2}} \int_{X^{+}_{m}} d\alpha^{+}_{m}(p)
\left[ e^{-i p\cdot x} a(p) + e^{i p\cdot x} a^{*}(p) \right],
\label{a}
\ee
\be
B(x) \equiv {1\over (2\pi)^{3/2}} \int_{X^{+}_{m}} d\alpha^{+}_{m}(p)
\left[ e^{-i p\cdot x} b(p) + e^{i p\cdot x} b^{*}(p) \right],
\label{b}
\ee
and
\be
\psi(x) \equiv {1\over (2\pi)^{3/2}} \int_{X^{+}_{m}} d\alpha^{+}_{m}(p)
\sum_{s=1}^{2}\left[ e^{-i p\cdot x} u_{s}(p) d_{s}(p) 
+ e^{i p\cdot x} u^{c}_{s}(p) d^{*}_{s}(p) \right]
\label{psi}
\ee
where 
$u_{s}(p)$
are two independent solutions of positive energy of the free Dirac equation
properly normalized. Here 
$X^{+}_{m}$
is the upper hyperboloid of mass $m$ and
$\alpha^{+}_{m}(p)$
is the Lorentz invariant measure defined on this Borel set. One can see that
the formal integration of the formula (\ref{super1}) gives:
\bea
Q \equiv \sum_{s=1}^{2}\int_{X^{+}_{m}} d\alpha^{+}_{m}(p) \{
- i [a(p) u^{c}_{s}(p) d^{*}_{s}(p) - a^{*}(p) u_{s}(p) d(p) ]
\nonumber \\ 
+ \gamma_{5} 
[ b(p) u^{c}_{s}(p) d^{*}_{s}(p) - b^{*}(p) u_{s}(p) d_{s}(p) ] \}
\label{supercharge}
\eea
which is a perfectly well defined expression acting in the Fock space and it
will be taken as a definition. It is elementary to obtain the following (anti)
commutation relations:
\bea
[Q, a(p)] = - i \sum_{s} u_{s}(p) d_{s}(p), \quad
[Q, a^{*}(p)] = - i \sum_{s} u^{c}_{s}(p) d^{*}_{s}(p),
\nonumber \\
~[Q, b(p)] = \gamma_{5} \sum_{s} u_{s}(p) d_{s}(p), \quad
[Q, a^{*}(p)] = \gamma_{5} \sum_{s} u^{c}_{s}(p) d^{*}_{s}(p),
\nonumber \\
\{Q, d^{*}_{s}(p) \} = i [ a^{*}(p) + i \gamma_{5} b^{*}(p) ] u_{s}(p), \quad
\{Q, d_{s}(p) \} = - i [ a(p) + i \gamma_{5} b(p) ] u_{s}(p).
\label{Q-com}
\eea
and 
\be
Q^{c} = Q.
\label{majorana}
\ee
\be
U_{a,A} Q = Q U_{a,A}.
\ee

We will need relations (\ref{Q-com}) in the coordinate space:
\bea
[Q, A(x)] = - i \psi(x), \quad
[Q, B(x)] = \gamma_{5} \psi(x), \quad
\{Q_{\alpha}, \psi_{\beta}(x) \} = 
\nonumber \\
- \partial_{\mu}A(x) (\gamma^{\mu} C)_{\alpha\beta}
- i \partial_{\mu}B(x) (\gamma_{5}\gamma^{\mu} C)_{\alpha\beta}
- im A(x) C_{\alpha\beta} + m B(x) (\gamma_{5} C)_{\alpha\beta}.
\eea

As we have said in the preceding Section, the Bogoliubov construction of the
perturbation series starts with the first order term
$T(x)$.
We have the following result:
\begin{prop}
Let us define the operator:
\bea
T(x) \equiv
c_{1} \left[ m :A(x)^{3}: + m :A(x) B(x)^{2}: 
+ :\bar{\psi}(x) \psi(x) A(x): - i :\bar{\psi}(x) \gamma_{5} \psi(x) B(x): 
\right] \nonumber \\
+ c_{2} \left[ m^{2} :A(x)^{2}: + m^{2} :B(x)^{2}: 
+ {1\over 2} m :\bar{\psi}(x) \psi(x): \right] \qquad
\label{inter}
\eea
and spinor operator:
\bea
T^{\mu}(x) \equiv c_{1} \left[ - i :A(x)^{2} \gamma^{\mu} \psi(x):
+ i :B(x)^{2} \gamma^{\mu} \psi(x):
+ 2 :A(x) B(x) \gamma_{5} \gamma^{\mu} \psi(x): \right]
\nonumber \\
+ c_{2} \left[ - im :A(x) \gamma^{\mu} \psi(x): + :B(x) \gamma^{\mu} \psi(x):
\right]
\label{inter-mu}
\eea

Then, the following relation is true:
\be
[ Q_{\alpha}, T(x) ] = i {\partial \over \partial x^{\mu}} T^{\mu}_{\alpha}(x).
\label{superinv1}
\ee

Moreover, the most general Wick polynomial of canonical dimension
$\leq 4$
verifying (\ref{superinv1}) is of the type (\ref{inter}). 
\end{prop}

As in the case of gauge theories, the relation (\ref{superinv1}) expresses the
invariance with respect to supersymmetric transformations of the interaction
Lagrangian in the formal adiabatic limit. In this particular case, the weak
adiabatic limit probably exists due to the fact that the masses of the model
are strictly positive. Moreover, the following relations are verified:
\begin{itemize}
\item
$SL(2,\C)$-covariance: for any
$A \in SL(2,\C)$
we have
\be
U_{a,A} T(x) U^{-1}_{a,A} = T(\delta(A)\cdot x+a), \quad
U_{a,A} T^{\mu}(x) U^{-1}_{a,A} = {\delta(A^{-1})^{\mu}}_{\rho}
T^{\rho}(\delta(A)\cdot x+a).
\ee
\item
Causality:
\be
\left[T(x), T(y)\right] = 0, \quad 
\left[T^{\mu}(x), T^{\rho}(y)\right] = 0, \quad 
\left[T^{\mu}(x), T(y)\right] = 0, \quad 
\forall x,y \in \R^{4} \quad s.t. \quad x \sim y.
\ee
\item
Unitarity: suppose that
$c_{1}, c_{2} \in \R$;
then:
\be
T(x)^{*} = T(x), \quad
T^{\mu}(x)^{c} = T^{\mu}(x).
\ee
\end{itemize}

Let us notice that there are no derivatives in the expression of the
interaction Lagrangian (\ref{inter}), so we can apply the procedure outlined in
the preceding Section.

\subsection{Second Order Chronological Product}. 

We consider a perturbation theory in the sense of Bogoliubov taking as the
interaction Lagrangian the expression (\ref{inter}) with
$c_{1} = 1, \quad c_{2} = 0$.

First, we define some distributions with causal support which will be needed in
the next proposition:
\be
D_{m,k}(x) \equiv [D^{(+)}_{m}(x)]^{k} + (-1)^{k-1} [D^{(-)}_{m}(x)]^{k}, 
\quad \forall k \in \N^{*}.
\ee

Next, we consider a canonical causal splitting
$$
D_{m,k}(x) = D^{adv}_{m,k}(x) - D^{ret}_{m,k}(x), \quad \forall k \in \N^{*}
$$
verifying Lorentz covariance and preserving the order of singularity.  By
definition, this canonical causal splitting is obtained using the central
decomposition formula of \cite{Sc1}. This is possible because all masses are
positive. The causal decomposition of
$D_{m,1}(x) = D_{m}(x)$
induces a similar splitting for the distribution
$$
S_{m}(x) = S^{adv}_{m}(x) - S^{ret}_{m}(x).
$$
We will denote the corresponding retarded, advanced and Feynman distributions
by:
$D^{F}_{m,k}(x)$
and
$S^{F}_{m}(x)$
respectively. Then we have:
\begin{prop}
The generic form of the second order chronological product is:
\be
T(x,y) = T^{c}(x,y) + \delta(x-y) N(x).
\label{T2}
\ee
where
\bea
T_{2}^{c}(x,y) \equiv :T(x) T(y): + 6m^{2} D_{m,3}^{F}(x-y) {\bf 1}
\nonumber \\
- 4 [ (\partial^{2} - m^{2}) D_{m,2}^{F}(x-y) ] :A(x) A(y):
- 4 [ (\partial^{2} - m^{2}) D_{m,2}^{F}(x-y) ] :B(x) B(y):
\nonumber \\
+ 4i :\overline{\psi}(x) [ \gamma\cdot\partial D^{F}_{m}(x-y) ] \psi(y):
\nonumber \\
+ 9m^{2} D_{m}^{F}(x-y) :A(x)^{2} A(y)^{2}: 
+  m^{2} D_{m}^{F}(x-y) :B(x)^{2} B(y)^{2}: 
\nonumber \\
+ 4m^{2} D_{m}^{F}(x-y) :A(x) B(x) A(y) B(y):
\nonumber \\
+ 4 :\overline{\psi}(x) S^{F}_{m}(x-y) \psi(y) A(x) A(y):
- 4 :\overline{\psi}(x) \gamma_{5} S^{F}_{m}(x-y) \gamma_{5} \psi(y) B(x) B(y):
\nonumber \\
- 3m^{2} D_{m}^{F}(x-y) [ :A(x)^{2} B(y)^{2}: + (x \leftrightarrow y)]
\nonumber \\
+ 3m D_{m}^{F}(x-y) [ :A(x)^{2} \overline{\psi}(y) \psi(y):
+ (x \leftrightarrow y)]
\nonumber \\
+  m D_{m}^{F}(x-y) [ :B(x)^{2} \overline{\psi}(y) \psi(y):
+ (x \leftrightarrow y)]
\nonumber \\
- 2im D_{m}^{F}(x-y) [ :A(x) B(x) \overline{\psi}(y) \gamma_{5} \psi(y):
+ (x \leftrightarrow y)]
\nonumber \\
- 4i [ :\overline{\psi}(x) S^{F}_{m}(x-y) \gamma_{5} \psi(y) A(x) B(y):
+ (x \leftrightarrow y)]
\eea
and
$N(x)$
is a finite normalization.
\end{prop}
                       
The proof consists in the  explicit computation of the commutator
$D_{2}$
like in \cite{standard}. The contribution
$T^{c}(x,y)$
correspond to the canonical causal splitting of the numerical distributions.
It was noticed from the very beginning \cite{WZ}, \cite{IZ} that the various
distributions appearing in the preceding formula have interesting properties:
for instance the distribution appearing as the coefficients of
$:A(x) A(y):$,
$:B(x) B(y):$,
$:\bar{\psi}(x) \psi(y):$
and
${\bf 1}$
are obtained from 
$D^{F}_{m,2}$
and
$D^{F}_{m,3}$
by simple operations. These properties can be preserved by the process of
distribution splitting. Moreover, the process of distribution splitting is
non-trivial only for
$D_{m,k}(x), \quad k = 2, 3$.
This corresponds to the assertion that one needs only two renormalization
constants for the Wess-Zumino model - see \cite{WZ} and \cite{IZ}.

Now we have:
\begin{thm}
In the conditions of the preceding proposition, the second order chronological
product 
$T(x,y)$
can be chosen such that it verifies:
\be
[ Q, T(x,y) ] = i{\partial \over \partial x^{\mu}} T^{\mu}_{1}(x,y) +
i{\partial \over \partial y^{\mu}} T^{\mu}_{2}(x,y)
\label{superinv2}
\ee
for some associated chronological products
$T^{\mu}_{i}(x,y), \quad i = 1,2$
if one takes in (\ref{T2}):
\be
N(x) \equiv i :A(x)^{4}: + i :B(x)^{4}: + 2 i :A(x)^{2} B(x)^{2}:  
\label{int2}
\ee
\label{super2}
\end{thm}

{\bf Proof:}
We follow the model of \cite{fermi} and compute the commutators:
\be
D^{\mu}_{1}(x,y) \equiv [T^{\mu}_{1}(x), T_{1}(y)], \quad
D^{\mu}_{2}(x,y) = D^{\mu}_{1}(y,x).
\label{D1mu}
\ee

By direct computation we have:
\bea
[T^{\mu}_{1}(x), T_{1}(y)] = 
- 2i :A(x)^{2} \gamma^{\mu} S_{m}(x-y) \psi(y) A(y):
- 2 :A(x)^{2} \gamma^{\mu} S_{m}(x-y) \gamma_{5} \psi(y) B(y):
\nonumber \\
+ 2i :B(x)^{2} \gamma^{\mu} S_{m}(x-y) \gamma_{5} \psi(y) B(y):
+ 2 :B(x)^{2} \gamma^{\mu} S_{m}(x-y) \psi(y) A(y):
\nonumber \\
+ 4 :A(x) B(x) \gamma_{5} \gamma^{\mu} S_{m}(x-y) \psi(y) A(y): \qquad
\nonumber \\
- 4i :A(x) B(x) \gamma_{5} \gamma^{\mu} S_{m}(x-y) \psi(y) B(y):
+ \cdots \qquad
\label{d2}
\eea
where the expressions 
$\cdots$
cannot produce anomalies.

We perform the canonical causal splitting of the expression
${\partial \over \partial x^{\mu}} D^{\mu}_{1}(x,y)$
and obtain the usual delta-distribution anomaly:
\bea
A_{1}(x,y) \equiv 2 \delta(x-y) \times
[ - i :A(x)^{3} \psi(x): + :B(x)^{3} \gamma_{5} \psi(x): 
\nonumber \\
+ :A(x)^{2} B(x)^{2} \gamma_{5}\psi(x): 
- i :A(x) B(x)^{2} \psi(x): ]
\label{ano2}
\eea
and a similar contribution follows from the other commutator.  But one easily
proves that:
\be
A_{1}(x,y) = {1\over2} [ Q, :A(x)^{4}: + :B(x)^{4}: + 2 :A(x)^{2} B(x)^{2}: ]
\ee
so the ``anomalies" can be eliminated by a proper choice of the finite
renormalization 
$N(x)$.
$\qed$

We remark that, quite similarly to the case of Yang-Mills theories, we have
obtained the second order contribution of the usual Wess-Zumino Lagrangian
from the formulation without the supplementary fields \cite{WZ}. However, in
this case, the anomalies can be completely eliminated by a proper choice of the
finite renormalization. Moreover, the arbitrariness of
$T(x,y)$
is of the form 
$\delta(x-y) \times (\ref{inter})$
if one requires that the canonical dimension does not exceeds $4$. This is
again consistent with the assertion from the traditional approaches to
renormalization theory. In this model, at least up to order $2$, one needs to
renormalization only two constants: the mass and the overall coupling constant.
\newpage
\section{Ward Identities and Anomalies\label{ward}}

\subsection{The Main Theorem}

We consider the Wess-Zumino model as defined by the Lagrangian (\ref{inter})
and show that we can implement supersymmetry invariance in all orders of
perturbation theory.
\begin{thm}
One can construct the chronological products
$T(X)$
such that, beside Bogoliubov axioms, the following relation is valid:
\be
[ Q, T(X) ] = i \sum_{l=1}^{n} {\partial \over \partial x^{\mu}_{l}} 
T^{\mu}_{l}(X), \quad \forall \quad |X|
\label{superinv}
\ee
where 
$T^{\mu}_{l}(X)$
are some auxiliary chronological products which can be chosen such that:
\be
T^{\mu}_{l}(X)^{c} = T^{\mu}_{l}(X), \quad \forall \quad |X|.
\label{c} 
\ee
\end{thm}

{\bf Proof:}

The main trick is to formulate carefully the {\bf induction hypothesis}. 
We suppose that we have constructed the chronological products 
$T(x_{1},\cdots,x_{p}), \quad p = 1, \dots, n - 1$
having the following properties: (\ref{sym})-(\ref{causality}) and
(\ref{unitarity}) for 
$|X| \leq n - 1$.
We also suppose that we have a more precise form of the Wick expansion
property:
\be
T(X) = \sum_{|I| = |J|} 
:\prod_{i \in I} \bar{\psi}_{\alpha_{i}}(x_{i}) 
t_{I,J,K,P}(X)_{\alpha_{I};\beta_{J}}
\prod_{j \in J} \psi_{\beta_{i}}(x_{j}) 
\prod_{k \in K} A(x_{k}) \prod_{p \in P} B(x_{p}):
\label{wick-wz}
\ee
where (a) the sum runs over all distinct triplets 
$I, J, K, P \subset \{1,\dots,n-1\}$;
(b) we have denoted
$\alpha_{I} \equiv \{\alpha_{i}\}_{i \in I}$
and
$\beta_{J} \equiv \{\beta_{j}\}_{j \in J}$;
(c) the expressions
$t_{I,J,K,P}(X)$
are numerical distributions (renormalized Feynman amplitudes); more precisely,
they take values in the matrix space
$M_{\C}(4,4)^{\otimes |I|}$;
(d) they are 
$SL(2,\C)$-covariant such that we have (\ref{invariance});
(e) we can suppose convenient (anti)-symmetry properties of the numerical
distributions without losing generality; 
(f) we have the limitation:
\be
\omega(t_{I,J,K,P}) \leq 4 - 3|I| - |K| - |L|.
\label{deg-t-wz}
\ee
We note that in this case we also have (\ref{commute}) for
$|X_{1}| + |X_{2}| \leq n - 1$.

We also suppose that we have constructed the Wick polynomials
$T^{\mu}_{l}(X), \quad |X \leq n - 1$
such that we have properties analogue to (\ref{sym}), (\ref{causality}),
and (\ref{wick-wz}). We use the convention:
\be
T(\emptyset) \equiv {\bf 1}, \quad 
T^{\mu}_{l}(\emptyset) \equiv 0, \quad
T^{\mu}_{l}(X) \equiv 0, \quad {\rm for} \quad l \not\in X.
\label{empty}
\ee

Then the induction hypothesis is supplemented as follows.
\begin{itemize}
\item
Symmetry:
\be
T_{\pi^{-1}(l)}^{\mu}(x_{\pi(1)},\cdots x_{\pi(p)}) = 
T_{l}^{\mu}(x_{1},\cdots x_{p}), 
\quad \forall \pi \in {\cal P}_{p}.
\label{sym-mu}
\ee
for
$p = 1, \dots, n - 1$;
\item
Covariance with respect to
$SL(2,\C)$:
\be
U_{a, A} T^{\mu}_{l}(x_{1},\cdots, x_{p}) U^{-1}_{a, A} =
{\delta(A^{-1})^{\mu}}_{\rho}
T^{\rho}_{l}(\delta(A)\cdot x_{1}+a,\cdots, \delta(A)\cdot x_{p}+a), 
\label{invariance-mu}
\ee
$p = 1, \dots, n - 1$;
\item
Charge conjugation invariance:
\be
U_{C} T^{\mu}_{l}(X) U^{-1}_{C} = T^{\mu}_{l}(X), \quad |X| \leq n - 1
\quad \Longleftrightarrow
T^{\mu}_{l}(X)^{c} = T^{\mu}_{l}(X), \quad |X| \leq n - 1
\label{charge-mu}
\ee
\item
Causality
\be
T^{\mu}_{l}(X_{1}X_{2}) = 
T^{\mu}_{l}(X_{1}) T(X_{2}) + T(X_{1}) T^{\mu}_{l}(X_{2})
\quad \forall X_{1} \geq X_{2}, \quad |X_{1}| + |X_{2}| \leq n - 1.
\label{causality-mu}
\ee
\item
Wick expansion property:
\be
T^{\mu}_{l}(X)_{\epsilon} = \sum_{|J| = |I| + 1} 
:\prod_{i \in I} \bar{\psi}_{\alpha_{i}}(x_{i}) 
t^{\mu}_{l;I,J,K,P}(X)_{\epsilon\alpha_{I};\beta_{J}}
\prod_{j \in J} \psi_{\beta_{i}}(x_{j})
\prod_{k \in K} A(x_{k}) \prod_{p \in P} B(x_{p}):
\label{wick-wz-mu}
\ee
where the sum runs over all distinct triplets 
$I, J, K, P \subset \{1,\dots,n-1\}$
verifying
$|J| = |I| + 1$;
the expressions
$t^{\mu}_{l;I,J,K,P}$
are numerical distributions taking values in the matrix space
$M_{\C}(4,4)^{\otimes |J|}$,
they are
$SL(2,\C)$-covariant
and with convenient (anti)-symmetry properties. Moreover, instead of
(\ref{deg-t}) we make the inductive hypothesis:
\be
\omega(t_{I,J,K,P}) \leq 3 - 3|I| - |K| - |L|.
\label{deg-t-mu}
\ee
We note that in this case we also have:
\be
[T^{\mu_{1}}_{l_{1}}(X_{1}), T^{\mu_{2}}_{l_{2}}(X_{2})] = 0, \quad
[T^{\mu}_{l}(X_{1}), T(X_{2})] = 0
\quad {\rm if} \quad X_{1} \sim X_{2}
\label{commute-mu}
\ee
for
$|X_{1}| + |X_{2}| \leq n - 1$.
\item
Supersymmetry invariance: we require that we have (\ref{superinv}) for
$|X| \leq n - 1$.
\item
In the case
$J = lJ'$
the distribution
$t^{\mu}_{l;I,J,K,P}(X)$
is ``proportional" to
$\gamma^{\mu}$
i.e. we have:
\be
t^{\mu}_{l;I,J,K,P}(X) = t_{l;I,J,K,P}(X) \otimes \gamma^{\mu}.
\label{1PI}
\ee

The corresponding Feynman graphs are 1-particle reducible.
\end{itemize}

We observe that the induction hypothesis is valid for 
$|X| = 1$
according to the preceding Section. We suppose that it is true for 
$|X| \leq n - 1$
and prove it for
$|X| = n$.

Now we can proceed in strict analogy with Subsection \ref{EG}. The proof of the
following items below goes in strict analogy to the proof of the similar
statements from the previous Subsection and can be easily provided with minimal
modifications. 

One constructs from
$T(X), \quad T^{\mu}_{l}(X), \quad |X| \leq n - 1$
the expressions
$\bar{T}(X), \quad |X| \leq n - 1$
according to (\ref{antichrono}) and similarly
$\bar{T}^{\mu}_{l}(X), \quad |X| \leq n - 1$
according to:
\bea
(-1)^{|X|} \bar{T}^{\mu}_{l}(X) \equiv \sum_{r=1}^{|X|} (-1)^{r} 
\sum_{X_{1},\cdots,X_{r} \in Part(X)}
[ T^{\mu}_{l}(X_{1}) T(X_{2}) \cdots T(X_{r}) + 
\cdots 
\nonumber \\
+ T(X_{1}) \cdots T(X_{r-1})T^{\mu}_{l}(X_{r})]; \quad
\label{antichrono-mu}
\eea
we use in an essential way the convention (\ref{empty}). 
Next, we define in analogy to (\ref{A1}) and (\ref{R1}) the expressions:
\be
A'^{\mu}_{l}(x_{1},\dots,x_{n-1};x_{n}) \equiv 
\sum_{\stackrel{X_{1},X_{2} \in Part(X)}{X_{2} \not= \emptyset, 
x_{n} \in X_{1}}}
\left[ T^{\mu}_{l}(X_{1}) \bar{T}(X_{2}) + T(X_{1}) \bar{T}^{\mu}_{l}(X_{2}) 
\right],
\label{A2}
\ee
\be
R'^{\mu}_{l}(x_{1},\dots,x_{n-1};x_{n}) \equiv 
\sum_{\stackrel{X_{1},X_{2} \in Part(X)}{X_{2} \not= \emptyset, 
x_{n} \in X_{1}}}
\left[ \bar{T}^{\mu}_{l}(X_{1}) T(X_{2}) + \bar{T}(X_{1}) T^{\mu}_{l}(X_{2}) 
\right].
\label{R2}
\ee

Next, we define in analogy to (\ref{D}) the expression
\be
D^{\mu}_{n}(x_{1},\dots,x_{n-1};x_{n}) \equiv 
A^{\prime\mu}_{l}(x_{1},\dots,x_{n-1};x_{n}) 
- R^{\prime\mu}_{l}(x_{1},\dots,x_{n-1};x_{n}).
\label{Dmu}
\ee
and prove that it has causal support i.e.
$supp(D^{\mu}_{n}(x_{1},\dots,x_{n-1};x_{n})) 
\subset \Gamma^{+}(x_{n}) \cup \Gamma^{-}(x_{n})$.
The proof is completely analogous to the standard proof from \cite{EG1}.

From the Wick expansion properties (\ref{wick-wz}) and (\ref{wick-wz-mu}) we
also have with the same conventions:
\be
D(X) = \sum_{|I| = |J|} 
:\prod_{i \in I} \bar{\psi}_{\alpha_{i}}(x_{i}) 
d_{I,J,K,P}(X)_{\alpha_{I};\beta_{J}}
\prod_{j \in J} \psi_{\beta_{i}}(x_{j}) 
\prod_{k \in K} A(x_{k}) \prod_{p \in P} B(x_{p}):
\label{wick-d-wz}
\ee
\be
D^{\mu}_{l}(X)_{\epsilon} = \sum_{|J| = |I| + 1} 
:\prod_{i \in I} \bar{\psi}_{\alpha_{i}}(x_{i}) 
d^{\mu}_{l;I,J,K,P}(X)_{\epsilon\alpha_{I};\beta_{J}}
\prod_{j \in J} \psi_{\beta_{i}}(x_{j})
\prod_{k \in K} A(x_{k}) \prod_{p \in P} B(x_{p}):
\label{wick-d-wz-mu}
\ee
where
$d^{\dots}_{\dots}(X)$
are numerical distributions verifying the following properties:
(a) $SL(2,\C)$-covariance;
(b) causal support i.e
$
supp(d^{\dots}_{\dots}(x_{1},\dots,x_{n-1};x_{n})) 
\subset \Gamma^{+}(x_{n}) \cup \Gamma^{-}(x_{n});
$
(c) limitation on the order of singularity:
\be
\omega(d_{I,J,K,P}) \leq 4 - 3|I| - |K| - |L|, \quad
\omega(d^{\mu}_{l;I,J,K,P}) \leq 3 - 3|I| - |K| - |L|.
\label{deg-d-mu}
\ee

The absence of derivative in the Wick monomials
$W_{i}(X)$
is again essential in establishing these relations.

As a consequence, there exists a 
$SL(2,\C)$-covariant causal splitting:
\be
D^{\mu}_{l}(x_{1},\dots,x_{n-1};x_{n}) =
A^{\mu}_{l}(x_{1},\dots,x_{n-1};x_{n}) - R^{\mu}_{l}(x_{1},\dots,x_{n-1};x_{n})
\label{decD-mu}
\ee
with
$supp(A^{\mu}_{l}(x_{1},\dots,x_{n-1};x_{n})) \subset \Gamma^{+}(x_{n})$
and
$supp(R^{\mu}_{l}(x_{1},\dots,x_{n-1};x_{n})) \subset \Gamma^{-}(x_{n})$
for all
$l = 1, \dots, n.$

We also have from the induction hypothesis in analogy with (\ref{conj}):
\be
D^{\mu}_{l}(x_{1},\dots,x_{n-1};x_{n})^{c} = 
(-1)^{n-1} D^{\mu}_{l}(x_{1},\dots,x_{n-1};x_{n}).
\label{conjD}
\ee

Now we investigate the possible obstruction to the extension of the
identity (\ref{superinv}) for
$|X| = n$.
We first prove by direct computation that we have:
\be
[ Q, D(X) ] = i \sum_{l \in X} {\partial \over \partial x^{\mu}_{l}} 
D^{\mu}_{l}(X), \quad |X| = n
\label{anoD}
\ee

We substitute here the causal decompositions (\ref{decD}) and (\ref{decD-mu})
in the preceding relation and we get:
\be
[ Q, A(X) ] - i \sum_{l=1}^{n} {\partial \over \partial x^{\mu}_{l}} 
A^{\mu}_{l}(X) \quad = \quad
[ Q, R(X) ] - i \sum_{l=1}^{n} {\partial \over \partial x^{\mu}_{l}} 
R^{\mu}_{l}(X).
\ee

Now the left hand side has support in 
$\Gamma^{+}(x_{n})$
and the right hand side in
$\Gamma^{-}(x_{n})$
so the common value, denoted by
$P(X)$
should have the support in
$\Gamma^{+}(x_{n}) \cap \Gamma^{-}(x_{n}) = \{x_{1} = \cdots = x_{n}\}$.
This means that we have:
\be
[ Q, A(X) ] - i \sum_{l=1}^{n} {\partial \over \partial x^{\mu}_{l}} 
A^{\mu}_{l}(X) = P(X)
\label{ano}
\ee
where
$P(X)$
has the structure:
\be
P(X) = \sum_{i} \left[p_{i}(\partial) \delta^{n-1}(X) \right] \quad W_{i}(x);
\label{wickP}
\ee
here 
$p_{i}$
are polynomials in the derivatives with the maximal degree restricted by
\be
deg(p_{i}) + deg(W_{i}) \leq 5
\label{degP}
\ee
and
\be
\delta^{n-1}(X) \equiv \delta(x_{1}-x_{n}) \cdots \delta(x_{n-1}-x_{n}).
\ee

It is easy to see that the ``anomaly" can be produced only by those terms
appearing in the Wick expansions of
$D(X)$
and
$D^{\mu}_{l}(X)$
for which the Wick monomials are restricted by
$\omega(W_{i}) \leq 5$.
We will show in the next Subsections that one can choose
$P(X) = 0$.
We will write such a generic form of these terms from
$D(X)$
and
$D^{\mu}_{l}(X)$
in the next two Subsections. 

\subsection{The expression of $D(X)$\label{D's}}

The terms corresponding to canonical dimension $\leq 5$ from (\ref{wick-d-wz})
are:
\vskip 0.5cm
\centerline{A. \underline{$\omega(W_{K}) = 1$}}
\vskip 0.5cm
\be
D^{(1)}(X) = \sum d^{(1)}_{i}(X) \quad A(x_{i})
\ee
\be
D^{(2)}(X) = \sum d^{(2)}_{i}(X) \quad B(x_{i})
\ee
\newpage
\vskip 1cm
\centerline{B. \underline{$\omega(W_{K}) = 2$}}
\vskip 1cm
\be
D^{(3)}(X) = \sum d^{(3)}_{ij}(X) \quad :A(x_{i}) A(x_{j}):
\ee
\be
D^{(4)}(X) = \sum d^{(4)}_{ij}(X) \quad :A(x_{i}) B(x_{j}):
\ee
\be
D^{(5)}(X) = \sum d^{(5)}_{ij}(X) \quad :B(x_{i}) B(x_{j}):
\ee
\vskip 1cm
\centerline{C. \underline{$\omega(W_{K}) = 3$}}
\vskip 1cm
\be
D^{(6)}(X) = \sum d^{(6)}_{ijk}(X) \quad :A(x_{i}) A(x_{j}) A(x_{k}):
\ee
\be
D^{(7)}(X) = \sum d^{(7)}_{ijk}(X) \quad :A(x_{i}) A(x_{j}) B(x_{k}):
\ee
\be
D^{(8)}(X) = \sum d^{(8)}_{ijk}(X) \quad :A(x_{i}) B(x_{j}) B(x_{k}):
\ee
\be
D^{(9)}(X) = \sum d^{(9)}_{ijk}(X) \quad :B(x_{i}) B(x_{j}) B(x_{k}):
\ee
\be
D^{(10)}(X) = \sum :\overline{\psi}(x_{i}) d^{(10)}_{ij}(X)  \psi(x_{j}):
\ee
\vskip 0.5cm
\centerline{D. \underline{$\omega(W_{K}) = 4$}}
\vskip 0.5cm
\be
D^{(11)}(X) = \sum d^{(11)}_{ijkp}(X) \quad 
:A(x_{i}) A(x_{j}) A(x_{k}) A(x_{p}):
\ee
\be
D^{(12)}(X) = \sum d^{(12)}_{ijkp}(X) \quad 
:A(x_{i}) A(x_{j}) A(x_{k}) B(x_{p}):
\ee
\be
D^{(13)}(X) = \sum d^{(13)}_{ijkp}(X) \quad 
:A(x_{i}) A(x_{j}) B(x_{k}) B(x_{p}):
\ee
\be
D^{(14)}(X) = \sum d^{(14)}_{ijkp}(X) \quad 
:A(x_{i}) B(x_{j}) B(x_{k}) B(x_{p}):
\ee
\be
D^{(15)}(X) = \sum d^{(15)}_{ijkp}(X) \quad 
:B(x_{i}) B(x_{j}) B(x_{k}) B(x_{p}):
\ee
\be
D^{(16)}(X) = \sum \quad 
:\overline{\psi}(x_{i}) d^{(16)}_{ijk}(X) \psi(x_{j}) A(x_{k}):
\ee
\be
D^{(17)}(X) = \sum \quad
:\overline{\psi}(x_{i}) d^{(17)}_{ijk}(X) \psi(x_{j}) B(x_{k}):
\ee
\newpage
\centerline{E. \underline{$\omega(W_{K}) = 5$}}
\vskip 1cm
\be
D^{(18)}(X) = \sum d^{(18)}_{ijkpq}(X) \quad 
:A(x_{i}) A(x_{j}) A(x_{k}) A(x_{p}) A(x_{q}):
\ee
\be
D^{(19)}(X) = \sum d^{(19)}_{ijkpq}(X) \quad 
:A(x_{i}) A(x_{j}) A(x_{k}) A(x_{p}) B(x_{q}):
\ee
\be
D^{(20)}(X) = \sum d^{(20)}_{ijkpq}(X) \quad 
:A(x_{i}) A(x_{j}) A(x_{k}) B(x_{p}) B(x_{q}):
\ee
\be
D^{(21)}(X) = \sum d^{(21)}_{ijkpq}(X) \quad 
:A(x_{i}) A(x_{j}) B(x_{k}) B(x_{p}) B(x_{q}):
\ee
\be
D^{(22)}(X) = \sum d^{(22)}_{ijkpq}(X) \quad 
:A(x_{i}) B(x_{j}) B(x_{k}) B(x_{p}) B(x_{q}):
\ee
\be
D^{(23)}(X) = \sum d^{(23)}_{ijkpq}(X) \quad 
:B(x_{i}) B(x_{j}) B(x_{k}) B(x_{p}) B(x_{q}):
\ee
\be
D^{(24)}(X) = \sum \quad
:\overline{\psi}(x_{i}) d^{(24)}_{ijkp}(X) \psi(x_{j}) A(x_{k}) A(x_{p}):
\ee
\be
D^{(25)}(X) = \sum \quad
:\overline{\psi}(x_{i}) d^{(25)}_{ijkp}(X) \psi(x_{j}) A(x_{k}) B(x_{p}):
\ee
\be
D^{(26)}(X) = \sum \quad
:\overline{\psi}(x_{i}) d^{(26)}_{ijkp}(X) \psi(x_{j}) B(x_{k}) B(x_{p}):
\ee

The term proportional to the identity operator
${\bf 1}$
is omitted because it does not contribute to (\ref{anoD}). We assume that the
$d^{(10)}, d^{(16)}, d^{(17)}, d^{(24)} - d^{(26)}$
are matrix-valued distributions; more precisely they have values in
$M_{\C}(4,4)$.
Moreover, it can be proved that these distribution can be chosen such that they
verify
\be
C^{-1} d^{(\alpha)}_{ij\dots}(X) C = 
- d^{(\alpha)}_{ji\dots}(\pi_{ij}(X))^{T}
\ee
without losing generality. The other expressions 
$d^{(\alpha)}$
are numerical distributions. The distributions
$d^{(\alpha)}, \quad \alpha = 1,\dots,26$
are
$SL(2,\C)$-covariant 
and have causal support.

\subsection{The expression of $D^{\mu}_{l}(X)$\label{Dmu's}}
The terms corresponding to canonical dimension $\leq 5$ from
(\ref{wick-d-wz-mu}) are:
\vskip 0.5cm
\centerline{A. \underline{$\omega(W_{K}) = 3/2$}}
\vskip 0.5cm
\be
D^{(1)\mu}_{l}(X) = \sum d^{(1)\mu}_{l;i}(X) \quad \psi(x_{i})
\ee
\vskip 0.5cm
\centerline{B. \underline{$\omega(W_{K}) = 5/2$}}
\vskip 1cm
\be
D^{(2)\mu}_{l}(X) = \sum d^{(2)\mu}_{l;ij}(X) \quad :\psi(x_{i}) A(x_{j}):
\ee
\be
D^{(3)\mu}_{l}(X) = \sum d^{(3)\mu}_{l;ij}(X) \quad :\psi(x_{i}) B(x_{j}):
\ee
\vskip 1cm
\centerline{C. \underline{$\omega(W_{K}) = 7/2$}}
\vskip 1cm
\be
D^{(4)\mu}_{l}(X) = \sum d^{(4)\mu}_{l;ijk}(X) \quad 
:\psi(x_{i}) A(x_{j}) A(x_{k}):
\ee
\be
D^{(5)\mu}_{l}(X) = \sum d^{(5)\mu}_{l;ijk}(X) \quad 
:\psi(x_{i}) A(x_{j}) B(x_{k}):
\ee
\be
D^{(6)\mu}_{l}(X) = \sum d^{(6)\mu}_{l;ijk}(X) \quad 
:\psi(x_{i}) B(x_{j}) B(x_{k}):
\ee
\vskip 0.5cm
\centerline{D. \underline{$\omega(W_{K}) = 9/2$}}
\be
D^{(7)\mu}_{l}(X) = \sum d^{(7)\mu}_{l;ijkp}(X) \quad 
:\psi(x_{i}) A(x_{j}) A(x_{k}) A(x_{p}):
\ee
\be
D^{(8)\mu}_{l}(X) = \sum d^{(8)\mu}_{l;ijkp}(X) \quad 
:\psi(x_{i}) A(x_{j}) A(x_{k}) B(x_{p}):
\ee
\be
D^{(9)\mu}_{l}(X) = \sum d^{(9)\mu}_{l;ijkp}(X) \quad 
:\psi(x_{i}) A(x_{j}) B(x_{k}) B(x_{p}):
\ee
\be
D^{(10)\mu}_{l}(X) = \sum d^{(10)\mu}_{l;ijkp}(X) \quad 
:\psi(x_{i}) B(x_{j}) B(x_{k}) B(x_{p}):
\ee
\be
D^{(11)\mu}_{l}(X) = \quad \sum 
:\overline{\psi(x_{i})} d^{(11)\mu}_{l;ijk}(X) \psi(x_{j}) \psi(x_{k}): 
\ee
\vskip 1cm
We assume that the are matrix-valued distributions:
$d^{(1)\mu}_{l;I} - d^{(10)\mu}_{l;I} \in M_{\C}(4,4)$
and
$d^{(11)\mu}_{l;ijk} \in M_{\C}(4,4)^{\otimes 2}$;
we can impose for these distributions the condition
\be
C^{-1} d^{(\alpha)}_{l;I}(X) C = - d^{(\alpha)}_{l;I}(X)^{T}, 
\quad \alpha = 1,\dots, 10
\ee
and
\be
C^{-1} d^{(11)}_{l;ijk}(X) C = - d^{(11)}_{l;jik}(\pi_{ij} (X))^{T}
\ee
without losing generality. The distributions
$d^{(\alpha)\mu}, \quad \alpha = 1,\dots,11$
are
$SL(2,\C)$-covariant 
and have causal support.

Moreover, we have from the induction hypothesis (\ref{1PI}) that
\be
d^{(\alpha)}_{l;lI} = d^{(\alpha)}_{l;lI} \gamma^{\mu}, \quad
\alpha = 1, \dots, 10
\label{gamma1}
\ee
where
$d^{(\alpha)}_{i;I}$
are numerical distribution, and
\be
d^{(11)}_{j;ijk} = d^{(11)}_{j;ijk} \otimes \gamma^{\mu}
\label{gamma2}
\ee
where
$d^{(11)}_{j;ijk}$
is a matrix-valued distribution, more precisely with values in
$M_{\C}(4,4)$.

\subsection{The Basic Equations\label{basic}}
The expression
$
i [ Q, D(X) ] + \sum_{l} {\partial\over \partial x^{\mu}_{l}} D^{\mu}_{l}(X)
$
is a Wick sum and the relevant contributions following from the preceding two
Subsections are:
\vskip 1cm
\noindent 1.1 The coefficient of the monomial
$\psi(x_{i})$:
\be
d^{(1)}_{i}(X) + i d^{(2)}_{i}(X) \gamma_{5} - im d^{(1)}_{i;i}(X) 
+ \sum_{l} {\partial\over \partial x^{\mu}_{l}} d^{(1)\mu}_{l;i}(X)
\ee
\vskip 1cm
\centerline{B. \underline{$\omega(W_{K}) = 5/2$}}
\vskip 1cm
\noindent 2.1 The coefficient of the monomial
$:\psi(x_{i}) A(x_{j}):$
\be
2 d^{(3)}_{ij}(X) + i d^{(4)}_{ij}(X) \gamma_{5}  
-2m d^{(10)}_{ji}(X') - im d^{(2)}_{i;ij}(X)
+ \sum_{l} {\partial\over \partial x^{\mu}_{l}} d^{(2)\mu}_{l;ij}(X)
\ee
2.2 The coefficient of the monomial
$:\psi(x_{i}) B(x_{j}):$
\be
d^{(4)}_{ij}(X) + 2i d^{(5)}_{ij}(X) \gamma_{5} 
- 2im \gamma_{5} d^{(10)}_{ji}(X') - im d^{(3)}_{i;ij}(X)
+ \sum_{l} {\partial\over \partial x^{\mu}_{l}} d^{(3)\mu}_{l;ij}(X)
\ee
\vskip 1cm
\centerline{C. \underline{$\omega(W_{K}) = 7/2$}}
\vskip 1cm
\noindent 3.1 The coefficient of the monomial
$:\psi(x_{i}) A(x_{j}) A(x_{k}):$
\be
3 d^{(6)}_{ijk}(X) + i d^{(7)}_{jki}(X') \gamma_{5}
- 2 m d^{(16)}_{jik}(X') - im d^{(4)}_{i;ijk}(X)
+ \sum_{l} {\partial\over \partial x^{\mu}_{l}} d^{(4)\mu}_{l;ijk}(X)
\ee
3.2 The coefficient of the monomial
$:\psi(x_{i}) A(x_{j}) B(x_{k}):$
\bea
2 d^{(7)}_{ijk}(X) + 2i d^{(8)}_{jki;abc}(X') \gamma_{5}
- 2im d^{(16)}_{kij}(X') \gamma_{5} - 2m d^{(17)}_{jik}(X')  
\nonumber \\
- im d^{(5)}_{i;ijk}(X)
+ \sum_{l} {\partial\over \partial x^{\mu}_{l}} d^{(5)\mu}_{l;ijk}(X)
\eea
3.3 The coefficient of the monomial
$:\psi(x_{i}) B(x_{j}) B(x_{k}):$
\be
d^{(8)}_{ijk}(X) + 3i d^{(9)}_{ijk}(X) \gamma_{5} 
- 2im d^{(17)}_{jik}(X') \gamma_{5} - im d^{(6)}_{i;ijk}(X) 
+ \sum_{l} {\partial\over \partial x^{\mu}_{l}} d^{(6)\mu}_{l;ijk}(X)
\ee
3.4 The coefficient of the monomial
$:\partial_{\mu}A(x_{j}) \psi(x_{j}):$
\be
2i \gamma^{\mu} d^{(10)}_{ij}(X) + d^{(2)\mu}_{i;ji}(X')
\ee
3.5 The coefficient of the monomial
$:\partial_{\mu}B(x_{i}) \psi(x_{j}):$
\be
- 2 \gamma_{5} \gamma^{\mu} d^{(10)}_{ij}(X) + d^{(3)\mu}_{i;ji}(X')
\ee

\vskip 1cm
\centerline{D. \underline{$\omega(W_{K}) = 9/2$}}
\vskip 1cm
\noindent 4.1 The coefficient of the monomial
$:\psi(x_{i}) A(x_{j}) A(x_{k}) A(x_{p}):$
\be
4 d^{(11)}_{ijkp}(X) + i d^{(12)}_{pijk}(X') \gamma_{5}
- 2 m d^{(24)}_{jikp}(X') - im d^{(7)}_{i;ijkp}(X)
+ \sum_{l} {\partial\over \partial x^{\mu}_{l}} d^{(7)\mu}_{l;ijkp}(X)
\ee
4.2 The coefficient of the monomial
$:\psi(x_{i}) A(x_{j}) A(x_{k}) B(x_{p}):$
\bea
3 d^{(12)}_{ijkp}(X) + 2i d^{(13)}_{kijp}(X') \gamma_{5}
- 2im d^{(24)}_{pijk}(X') \gamma_{5} - 2m d^{(25)}_{jikp}(X') 
\nonumber \\
- im d^{(8)}_{i;ijkp}(X)
+ \sum_{l} {\partial\over \partial x^{\mu}_{l}} d^{(8)\mu}_{l;ijkp}(X)
\eea
4.3 The coefficient of the monomial
$:\psi(x_{i}) A(x_{j}) B(x_{k}) B(x_{p}):$
\bea
2 d^{(13)}_{ijkp}(X) + 3i d^{(14)}_{jikp}(X') \gamma_{5}
- 2im d^{(25)}_{kij}(X') \gamma_{5} - 2m d^{(26)}_{jikp}(X') 
\nonumber \\
- im d^{(9)}_{i;ijkp}(X)
+ \sum_{l} {\partial\over \partial x^{\mu}_{l}} d^{(9)\mu}_{l;ijkp}(X)
\eea
4.4 The coefficient of the monomial
$:\psi(x_{i}) B(x_{j}) B(x_{k}) B(x_{p}):$
\be
d^{(14)}_{ijkp}(X) + 4i d^{(15)}_{ijkp}(X) \gamma_{5}
- 2im d^{(26)}_{jikp}(X') \gamma_{5} - im d^{(10)}_{i;ijkp}(X)
+ \sum_{l} {\partial\over \partial x^{\mu}_{l}} d^{(10)\mu}_{l;ijkp}(X)
\ee
4.5 The coefficient of the monomial
$:\psi(x_{i}) \partial_{\mu}A(x_{j}) A(x_{k}):$
\be
2i \gamma^{\mu} d^{(16)}_{jik}(X') + d^{(4)\mu}_{j;ijk}(X) 
\ee
4.6 The coefficient of the monomial
$:\psi(x_{i}) \partial_{\mu}A(x_{j}) B(x_{k}):$
\be
2i \gamma^{\mu} d^{(17)\mu}_{jik}(X') + d^{(5)\mu}_{j;ijk}(X)
\ee
4.7 The coefficient of the monomial
$:\psi(x_{i}) A(x_{j}) \partial_{\mu}B(x_{k}):$
\be
- 2 \gamma_{5} \gamma^{\mu} d^{(16)}_{kij}(X') + d^{(5)\mu}_{k;ijk}(X) 
\ee
4.8 The coefficient of the monomial
$:\psi(x_{i}) \partial_{\mu}B(x_{j}) B(x_{k}):$
\be
- \gamma_{5} \gamma^{\mu} d^{(17)}_{jik}(X') + d^{(6)\mu}_{j;ijk}(X)
\ee
4.9 The coefficient of the monomial
$:\overline{\psi}(x_{i}) \psi(x_{j}) \psi(x_{k}):$
\be
d^{(16)}_{ijk}(X) \otimes {\bf 1} + i d^{(17)}_{ijk}(X) \otimes \gamma_{5} 
- 3im d^{(11)}_{k;ijk}(X)
+ \sum_{l} {\partial\over \partial x^{\mu}_{l}} d^{(11)\mu}_{l;ijk}(X) 
\ee

In these equation we mean by 
$X'$
the corresponding permutation of the variables $X$.
The expression of the anomaly
$P(X)$
is given by (\ref{wickP}) where only the 17 Wick monomials listed above can
appear.

All anomalies
$p_{i}$
can be eliminated purely algebraically by redefining some of the causal
splittings. First, we take a causal splitting of all distributions
\be
d^{(\alpha)} = a^{(\alpha)} - r^{(\alpha)},
\quad
d^{(\alpha)\mu} = a^{(\alpha)\mu} - r^{(\alpha)\mu}
\ee
verifying
$SL(2,\C)$-covariance 
and preserving the order of singularity. (We make the labelling in such a way
that for
$\alpha = 1,\dots,26$
and
$\alpha = 1,\dots,11$
respectively we have the distributions from the preceding two Subsections.)
The expressions
$A(X)$
and
$A^{\mu}_{l}(X)$
are defined according to the relations of the type (\ref{wick}).

Then we notice that we can absorb all anomalies in:
\bea
a^{(1)}_{i}(X), a^{(3)}_{ij}(X), a^{(4)}_{ij}(X), a^{(6)}_{ijk}(X),
a^{(7)}_{ijk}(X), a^{(8)}_{ijk}(X), a^{(2)\mu}_{j;ji}(X), 
a^{(3)\mu}_{j;ji}(X), a^{(11)}_{ijkp}(X), 
\nonumber \\
a^{(12)}_{ijkp}(X), a^{(13)}_{ijkp}(X), a^{(14)}_{ijkp}(X), 
a^{(4)\mu}_{j;ijk}(X), a^{(5)\mu}_{j;ijk}(X), a^{(5)\mu}_{k;ijk}(X),
a^{(6)\mu}_{j;ijk}(X), a^{(16)}_{ijk}(X)
\eea
respectively.

This means that one can make the causal splitting such that (\ref{ano}) is:
\be
[ Q, A(X) ] = i \sum_{l=1}^{n} {\partial \over \partial x^{\mu}_{l}} 
A^{\mu}_{l}(X) 
\label{no-ano}
\ee

From the relation (\ref{no-ano}) one can obtain by Hermitian conjugation:
\be
[ Q, A(X)^{*} ] = i \sum_{l=1}^{n} {\partial \over \partial x^{\mu}_{l}} 
A^{\mu}_{l}(X)^{c} 
\label{no-ano-c}
\ee
where use has been made of the relation (\ref{majorana}) - which says that the
supercharge is a Majorana spinor - and of the relations (\ref{conj}) and
(\ref{conjD}).
If one makes the substitutions (\ref{redef}) and
\be
A^{\mu}_{l}(X) \rightarrow 
{1\over 2} \left[ A^{\mu}_{l}(X) + (-1)^{n-1} (A(X)^{\mu}_{l})^{c}\right]
\ee
we still have the a legitimate causal decomposition of the type (\ref{decD})
and (\ref{decD-mu}); the consistency is ensured by (\ref{conjD}).  It follows
that the causal splitting can be chosen such that
\be
A(X)^{*} = (-1)^{n-1} A(X), \quad
A^{\mu}_{l}(X)^{c} = (-1)^{n-1} A^{\mu}_{l}(X).
\label{conjA}
\ee

This relation is essential in establishing the unitarity axiom in order $n$
(see \cite{EG1}).

Finally we define the chronological products as in Subsection \ref{EG} and
by analogy
\be
T^{\mu}_{l}(X) \equiv A^{\mu}_{l}(X) - A'^{\mu}_{l}(X) \quad = \quad
R^{\mu}_{l}(X) - R'^{\mu}_{l}(X).
\ee

These expressions satisfy the Poincar\'e covariance, causality and unitarity
conditions. If we make the substitution (\ref{symmetrization}) and analogously:
\be
T^{\mu}_{l}(x_{1},\cdots, x_{n}) \rightarrow {1 \over n!}
\sum_{\pi} T^{\mu}_{\pi^{-1}(l)}(x_{\pi(1)},\cdots, x_{\pi(n)})
\ee
then we also have the symmetry axioms (\ref{sym}) and (\ref{sym-mu}). Finally,
one can prove that the rest of the induction hypothesis is true in order $n$ of
the perturbation theory. Some effort is required for (\ref{1PI}).

The invariance of the $S$-matrix with respect to space-time inversions can be
obtained as in the case of quantum electrodynamics \cite{Sc1}.
\section {Conclusions}

We have proved that the essence of the improved renormalizability properties of
supersymmetric models is due to the fact that the equation (\ref{anoD}) is of
purely algebraic nature and so the possible anomalies can be eliminated by a
redefinition of the causal splitting. We comment on the corresponding Ward
identities following from (\ref{superinv}). If one considers chronological
products of Wick submonomials with a proper normalization then one can
translate the equation (\ref{superinv}) into equations on the renormalized
Feynman amplitudes. One obtains that all expressions from Subsection
\ref{Dmu's} with
$d^{\dots}_{\dots} \rightarrow t^{\dots}_{\dots}$
are null. These are exactly the Ward identities of the Wess-Zumino model
\cite{IZ}. Such type of identities have been extensively studied in \cite{Du2}.
In particular, they impose the behaviour of the Feynman amplitudes described
before the theorem \ref{super2}.

A very interesting subject for further investigations is to determined how
general is the phenomenon exhibited in this paper, that is purely algebraic
Ward identities. 

A reformulation of the preceding analysis in terms of superfields \cite{SS},
\cite{WB}, \cite{We} would also be interesting.

\end{document}